\documentclass[11pt,twoside]{article}
\usepackage{asp2010}
\resetcounters

\begin{document}

\title{Identifying Variations to the IMF at High-$z$ Through Deep Radio Surveys}  
\author{Eric J. Murphy
\affil{{\it Spitzer} Science Center, MC 314-6, California Institute of Technology, Pasadena, CA 91125; emurphy@ipac.caltech.edu}
}

\begin{abstract}
In this article I briefly describe how deep radio surveys may provide a means to identify variations in the upper end of the initial mass function (IMF) in star-forming galaxies at high redshifts (i.e., $z\gtrsim$3).  
At such high redshifts, I argue that deep radio continuum observations at frequencies $\gtrsim$10 GHz using next generation facilities (e.g., EVLA, MeerKAT, SKA/NAA) will likely provide the most accurate measurements for the ionizing photon rates (star formation rates; SFRs) of normal galaxies since their non-thermal emission should be highly  suppressed due to the increased inverse Compton (IC) losses from the cosmic microwave background (CMB), leaving only thermal (free-free) emission detectable.  
Thus, a careful analysis of such observations in combination with future ALMA and JWST data, measuring the rest-frame far-infrared and UV emission from the same population of galaxies, may yield the best means to search for variability in the stellar IMF at such epochs.  
\end{abstract}

\section{Introduction}
While star-forming galaxies are currently thought to be responsible for completely reionizing the intergalactic medium (IGM) by $z\sim6$ \citep[e.g.][]{rhb01, fck06}, 
the ionizing flux arising from star formation in Lyman break galaxies at similar redshifts \citep{rjb07,rjb08} appears to fall a factor of $\ga$6 below the minimum value required to maintain an ionized IGM for a given clumping factor and escape fraction under the assumption of a Salpeter stellar initial mass function (IMF). 
\citet{rrc08} has shown that this discrepancy can be reconciled by flattening the stellar IMF to have a slope of $\sim-$1.7 if reionization occurred at $z = 9$ thus increasing the ionizing photon rate.  
While the idea of a top heavy IMF at these epochs does not seem completely inappropriate given that low metallicity environments will favor the production of more high-mass stars, trying to identify such a ariation remains difficult.  
Here, I describe a relatively simple test using existing data of star-forming regions in the nearby galaxy NGC~6946.  

\section{Data and SFR Calibrations}
A multi-wavelength star formation rate (SFR) comparison is presented for 10 star-forming regions in the nearby ($d = 6.8$~Mpc) galaxy NGC~6949, including its mildly starbursting nucleus \citep{ball85}.  
These regions were selected due their existing and forthcoming mid- and far-infrared spectroscopic data collected as part of the {\it Spitzer} Infrared Nearby Galaxies Survey \citep[SINGS;][]{rck03} and the project Key Insights on Nearby Galaxies: a Far-Infrared Survey with {\it Herschel} (KINGFISH; PI. R. Kennicutt).  
For a much more in depth description of this analysis the reader is referred to \cite{ejm10} and E.J. Murphy et al. (2010, in preparation).  

\subsection{Radio Data}
Radio imaging at 1.4~GHz ($14\arcsec \times12\farcs5$ beam) comes from the Westerbork Synthesis Radio Telescope (WSRT)-SINGS survey \citep{rb07}.  
Imaging at 1.5, 1.7, 4.9, and 8.5~GHz ($15\arcsec \times 15\arcsec$ beam) all come from \citet{beck07}.  
The 4.9 and 8.5~GHz radio data included single-dish measurements for short spacing corrections.  
Observations in the Ka-band ($26-40$) were taken using the Caltech Continuum Backend (CCB) on the GBT which simultaneously measures the entire Ka bandwidth over 4 equally spaced frequency channels.  
Reference beams are measured by nodding 1\farcm3 away from the source; details about the reference beam locations can ben found in Figure 1 of \citet{ejm10}.  
The average FWHM of the GBT beam in the Ka-band was $\approx$25\arcsec among our sets of observations, which projects to a physical scale of $\approx0.8$~kpc at the distance of NGC~6946.  
A detailed description on the performance of the CCB receiver, the data reduction pipeline, and error estimates, can be found in \citet{bm09}.

\begin{figure}
\begin{center}
\plotone{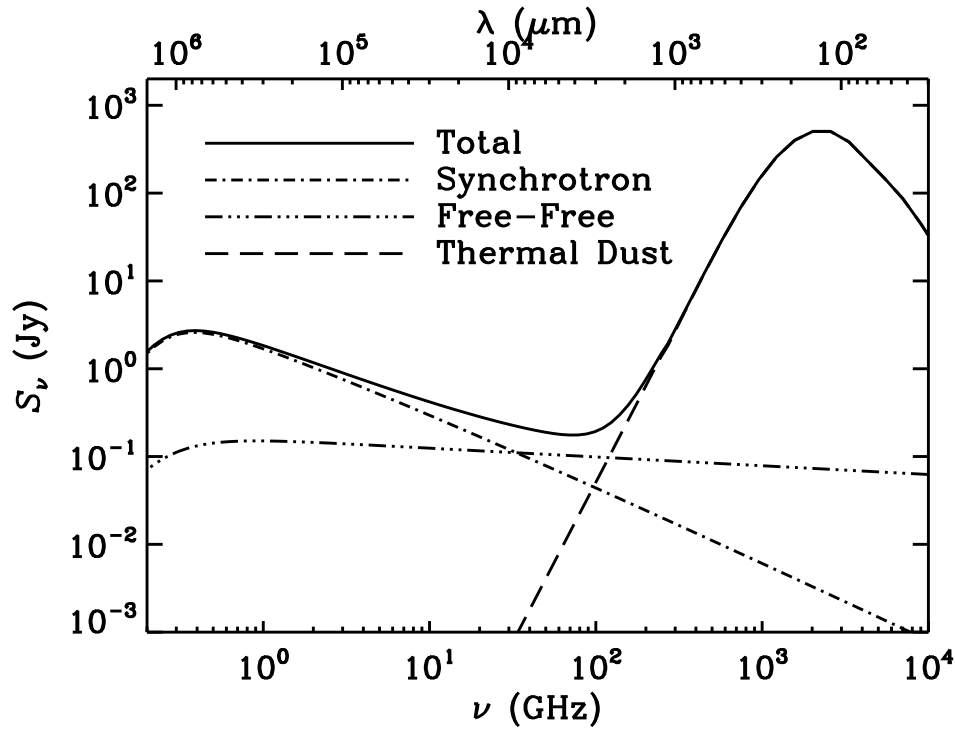}
\end{center}
\caption{A model radio to far-infrared galaxy spectrum \citep[see][]{ejm09c}.  
The individual contributions from non-thermal synchrotron, free-free, and thermal dust emission are indicated by dot-dashed, triple dot-dashed, and dashed lines, respectively.    
\label{fig:1}}
\end{figure}

\subsection{IR, UV, and Optical Data}
The 8, 24, and 850~$\mu$m data used here were included in the SINGS fifth data release.  
{\it Spitzer} 70~$\mu$m and SCUBA 450~$\mu$m data, also included in the SINGS fifth data release, are not used in lieu of new {\it Herschel} data taken as part of KINGFISH.  
Only KINGFISH data which have a resolution better than the 25" beam of our GBT Ka-band data are used, which includes PACS 70, 100, and 160~$\mu$m imaging, along with SPIRE 250 and 350~$\mu$m imaging.  
The infrared photometry are fit by the SED models of \citet{dh02} to derive total infrared (IR: $8-1000~\mu$m) luminosities from each region.  
{\it GALEX} far-UV (FUV; 1528~\AA) and near-UV (NUV; 2271~\AA) imaging were taken from the {\it GALEX} archive.  
H$\alpha$ imaging of NGC~6946 was taken from \citet{af98}.

\section{Calibration of SFR Diagnostics: Physical Assumptions}
To ensure that each SFR diagnostic can be compared fairly (i.e., borne from a common IMF), corresponding calibrations are computed using Starburst99 stellar population models \citep[][]{cl99}.  
I choose a Kroupa IMF \citep{pk01}, having a slope of -1.3 for stellar masses between $0.1-0.5~M_{\sun}$ and -2.3 for stellar masses ranging between $0.5-100~M_{\sun}$.  
A supernova cut-off mass of 8~$M_{\sun}$ is also assumed.  
This exercise is then repeated by flattening the slope of the upper mass function from -2.3 to -1.5.  
For both IMF cases, a solar metallicity and continuous star formation (i.e., a fixed SFR~$= 1M_{\sun}$~yr$^{-1}$) are assumed.  
The resultant Starburst99 output relevant for calibrating SFR diagnostics described below for both IMFs at an age of 100~Myr are given in Table \ref{murphy_e_tab1}.  

\begin{table}[!ht]
\caption{Starburst99 Output for Kroupa and Top Heavy IMFs \label{murphy_e_tab1}}
\smallskip
\begin{center}
{\small
\begin{tabular}{cccccc}
\tableline
\noalign{\smallskip}
IMF & $Q({\rm H}^{0})$ & $q_{\rm SNR}$ & $L_{\rm FUV}$ & $L_{\rm H\alpha}$ & $L_{\rm IR}$\\
&(photons~s$^{-1}$) &(SN~century$^{-1}$)&(ergs~s$^{-1}$~Hz$^{-1}$)& (ergs~s$^{-1}$)&(ergs~$s^{-1}$)\\
\noalign{\smallskip}
\tableline
\noalign{\smallskip}
Kroupa (-1.3,-2.3) & $1.37\times10^{53}$ & 1.16 & $1.14\times10^{28}$ & $1.86\times10^{41}$ & $2.58\times10^43$\\
Top Heavy (-1.3,-1.5) & $7.78\times10^{53}$ & 2.69 & $3.57\times10^{28}$ & $1.06\times10^{42}$ & $8.36\times10^43$\\
\tableline
 Top Heavy/Kroupa & 5.58 & 2.32 & 3.13 & 5.69 & 3.24\\
\noalign{\smallskip}
\tableline
\end{tabular}
}
\end{center}
Above Starburst99 output is taken after 100~Myr of continuous star
formation (1~$M_{\sun}$~yr$^{-1}$) assuming solar metallicity.  
\end{table}

\subsection{Radio SFRs}
Radio continuum emission from galaxies is generally composed of two optically thin components: non-thermal synchrotron emission associated with cosmic-ray (CR) electrons accelerated in a galaxy's magnetic field, and thermal bremsstrahlung (free-free) emission around massive star formation regions.  
The origin of each of these components lies in the process of massive star formation, which is most likely the physical underpinning of the nearly ubiquitous far-infrared (FIR)-radio correlation \citep[e.g.][]{gxh85,jc92,yrc01,ejm06a,ejm08}.  
In this simplistic model, the non-thermal radio luminosity of a galaxy at frequency $\nu$ is related to the SFR by supernova (SN) rate, $q_{\rm SNR}$, while the free-free radio luminosity is directly proportional to the number of ionizing photons, $Q({\rm H}^{0})$, along with a kinetic temperature ($T_{\rm e}$) dependence,  such that,
\begin{equation}
{\rm SFR}_{\nu-{\rm NT}} \propto q_{\rm SNR} \propto \nu^{\alpha} L_{\nu}^{\rm NT};\hspace {12pt}  
{\rm SFR}_{\nu-{\rm T}} \propto Q({\rm H}^{0}) \propto \nu^{0.1}T_{\rm e}^{-0.45}L_{\nu}^{\rm T}, 
\end{equation}
respectively, where $\alpha$ is the non-thermal spectral index.    
Thus, the SFR derived from the total radio emission at any one frequency can be described as a combination of each term such that, 
\begin{equation}
{\rm SFR}_{\nu} \propto L_{\nu}/(\nu^{-\alpha} + \nu^{-0.1}T_{\rm e}^{0.45}).  
\end{equation}

\subsection{IR, UV, H$\alpha$, and Hybrid SFRs}
Similarly, SFRs can be expressed using the output of Starburst99 in terms of IR, UV, and optical (H$\alpha$) observations, in a much more straightforward manner.  
IR, UV, and H$\alpha$ SFR estimates go as,
\begin{equation}
{\rm SFR}_{\rm IR} \propto L_{\rm IR};\hspace {12pt}  
 {\rm SFR}_{\rm UV} \propto L_{\rm UV};\hspace {12pt}  
{\rm SFR}_{{\rm H}\alpha} \propto Q({\rm H}^{0}) \propto L_{{\rm H}\alpha}.  
\end{equation} 
Since UV and optical wavelength suffer significantly from extinction, making it difficult to properly convert such measurements into an accurate SFR, a number of hybrid methods have been introduced into the literature.    
The simplest is using a combination of IR and (observed) UV SFR estimates to account for both the obscured and unobscured components, resulting in a `bolometric' or `total' SFR.  
Another hybrid diagnostic, which probably has a more solid physical backing, is the linear combination of observed H$\alpha$ and 24~$\mu$m observations to obtain an extinction corrected SFR \citep[e.g.,][]{rck07,dc07,rck09,dc10}; this diagnostic has been empirically calibrated against extinction corrected NIR recombination lines.   
Accordingly, these hybrid SFRs have the following observational dependencies: 
\begin{equation}
{\rm SFR}_{\rm TOT} \propto L_{\rm IR} + L_{\rm UV}; \hspace {12pt}  
{\rm SFR_{mix}} \propto Q({\rm H}^{0}) \propto L_{{\rm H}\alpha} + b\times  L_{24~\mu m},
\end{equation}
where $b$ is an empirically derived coefficient \citep[for a detailed discussion see][]{dc10}.

\begin{figure}
\plotone{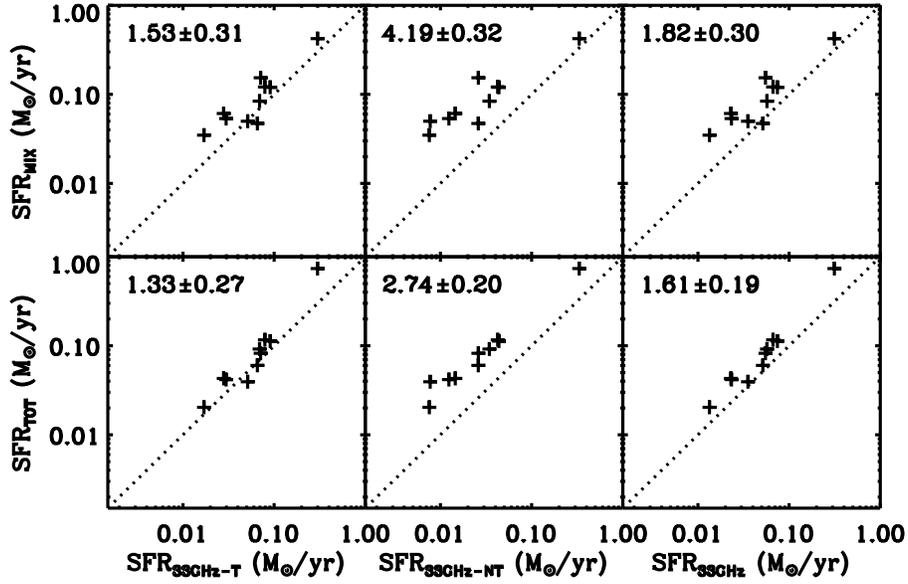}
\caption{A comparison of SFR diagnostics discussed in $\S$3.  
From left to right, the top row shows the differences between the H$\alpha + $24~$\mu$m and the 33~GHz thermal, non-thermal, and total SFR estimates.  
The same is shown in the bottom row, except the 33~GHz SFRs are compared to the IR$+$UV SFRs.  
In the top left corner of each panel, the median of the ordinate/abcissa ratio values, along with the dispersion about the fit to each trend are given.  
\label{fig:2}}
\end{figure}

\begin{figure}
\plotone{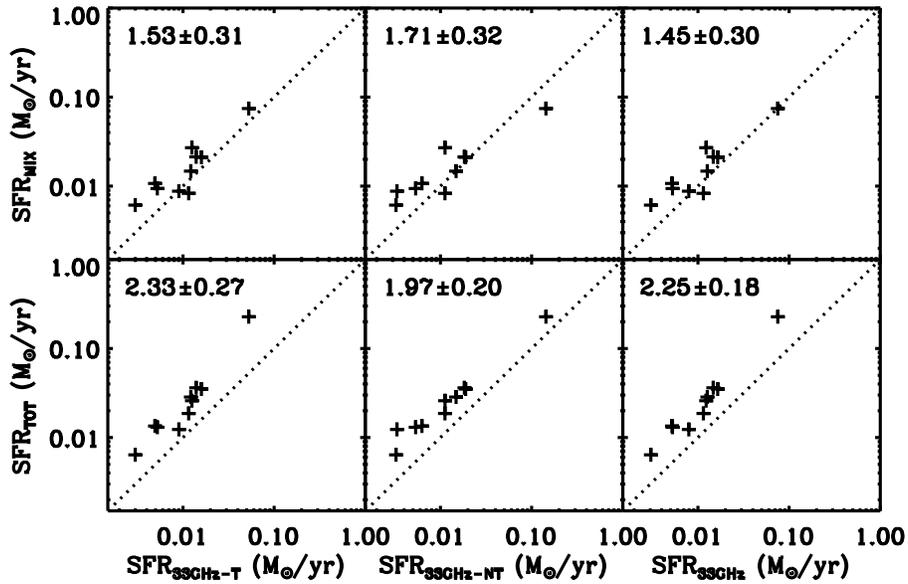}
\caption{Same as Figure \ref{fig:2}, except SFR calibrations have been derived using a top-heavy IMF (see Table \ref{murphy_e_tab1}).  
\label{fig:3}}
\end{figure}

\section{Comparison of SFR Diagnostics}
In Figure \ref{fig:2} SFRs from each region in NGC~6946 are plotted using the various calibrations discussed in $\S$3:  
H$\alpha + $ 24~$\mu$m SFRs are plotted against the thermal, non-thermal, and total 33~GHz SFR calibrations in the top row of panels, while the IR$+$UV SFRs are plotted against the same 33~GHz SFR calibrations in the bottom row of panels.  
I assume that the thermal radio continuum at 33~GHz provides the most accurate SFR estimate among all diagnostics as it is the most straightforward to relate to the ionizing photon rate (assuming that the spectral decomposition into thermal and non-thermal components is done correctly).  
SFR estimates using the thermal radio continuum at 33~GHz and the combination of H$\alpha +$24~$\mu$m emission appear to agree quite well over then entire luminosity range, having a median near unity, albeit, with some scatter.  
The non-thermal 33~GHz SFRs appear to underestimate the actual SFR by a factor of $\sim$2, on average, while showing good agreement in the nucleus.  
Also notable is the increased scatter among the extranuclear H{\sc ii} regions.  
This result can be attributed to the fact that the population of CR electrons in the extranuclear star-forming regions may be typically older, on average, than those being freshly injected into the ISM near  the starbusting nucleus.  
Thus, they have had time to propagate much larger distances, and no longer contribute to the non-thermal emission in the immediate vicinity of the H{\sc ii} regions \citep{ejm06b,ejm08}.      
On the other hand, the SFR estimated using the total radio luminosity (i.e., the combination of the free-free and non-thermal emission as given in Equation 2) shows good agreement with the H$\alpha +$24~$\mu$m estimate.  
This is due large thermal fractions at 33~GHz.  
The IR$+$UV SFR estimate also appears to do generally well relative to the 33~GHz thermal radio continuum, save the nucleus, which has IR$+$UV SFR which is $\sim$2.5 times larger.

In Figure \ref{fig:3} the same comparison of SFR diagnostics is shown as in Figure \ref{fig:2}, except that the top heave IMF calibrations have been used (see Table \ref{murphy_e_tab1}).  
The Starburst99 output suggests that the ionizing flux will increase by a factor of $\approx$6 by moving to a top heavy IMF, while the UV and IR output will only increase by a factor of $\approx$3.   
One of course expects that SFRs which depend only on the ionizing flux should be similar independent of the IMF.  
However, based on these differences in the calibrations, SFRs proportional to the ionizing flux should be a factor of $\sim$2 larger those which depend on the UV and/or IR luminosities for a top heavy IMF compared to a standard Kroupa IMF.  
In looking at the top and bottom left panels, this is observed;  
the thermal radio continuum and H$\alpha + $24~$\mu$m SFRs diagnostics still yield a nearly one-to-one correlation with one another while the IR$+$UV SFRs yield a SFR which is a factor of $\sim$2 discrepant from the 33~GHz thermal radio continuum estimate (i.e., a factor of $\sim$2 larger since the IMF of NGC~6946 more like a Kroupa IMF).  
Similarly, the same factor of $\sim$2 discrepancy is found when comparing the IR$+$UV SFRs to the SFRs estimated from the total radio continuum.  
I note that all of these three quantities used to estimate the SFRs in this case are as observed; no extinction correction was necessary, nor does it depend on an empirical calibration.  
Thus, by having UV, IR, and thermal radio continuum emission in hand, one should be able to quantify such discrepancies between SFR diagnostics. and search for sources with potentially top heavy IMFs.

\section{Radio Emission from Galaxies at High-{\it z}}
For a typical star-forming galaxy, CR electrons primarily lose their energy due to synchrotron and inverse Compton (IC) processes \citep[e.g.,][]{jc92}, although escape also plays a role \citep[e.g.,][]{hb93}.  
At $z = 0$, $U_{\rm CMB}\sim 4.2 \times 10^{-13} ~{\rm erg~cm}^{-3}$, which is significantly smaller than the radiation field energy density of the Milky Way (i.e., $U_{\rm MW} \sim 10^{-12} ~{\rm erg~ cm}^{-3}$).  
Thus, CR electron energy losses from IC scattering off the CMB are negligible at low redshifts. However, $U_{\rm CMB} \propto (1 + z)^{4}$, making such losses increasingly important with redshift.  
As an example, by $z\sim3$, $U_{\rm CMB}\sim1.1\times10^{-10}~$erg~cm$^{-3}$; 
equating this to the magnetic field energy density $U_{B} = B^{2}/(8\pi)$ results in a corresponding magnetic field strength of $\sim$50$\mu$G.  
This is nearly an order of magnitude larger than the ambient field strength in the solar neighborhood.  
Consequently, the non-thermal component of a galaxy's radio continuum emission will be increasingly suppressed with increasing redshift, eventually resulting in only the thermal component being detectable.  

\begin{figure}
\begin{center}
\plotone{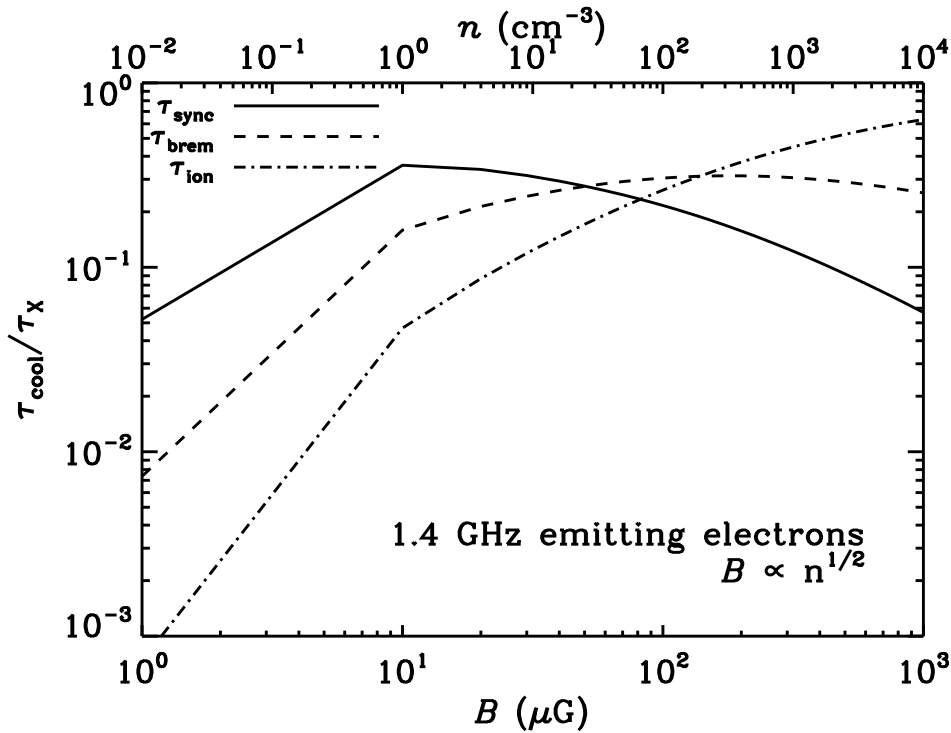}
\end{center}
\caption{The relative cooling fractions for synchrotron, bremsstrahlung, and ionization processes for 1.4~GHz emitting CR electrons as a function of increasing magnetic field strength under the assumptions that 
$(\nu/{\rm GHz)} = 1.3\times10^{-2} (B/\mu{\rm G})(E/{\rm GeV})^{2}$,  
$(n_{\rm ISM}/{\rm cm^{-3}}) \sim (B/10~\mu{\rm G})^{2}$, and 
$U_{\rm rad} \sim U_{B}$ (i.e., internal IC and synchrotron cooling timescales are similar).  
\label{fig:4}}
\end{figure}

It is worth noting that additional energy-loss terms may become increasingly important in the case of galaxies hosting strong starbursts.   
In such systems, whose energetics and ISM may be vastly different, ionization, bremsstrahlung, and adiabatic cooling through advection out of a galaxy by galactic scale winds can all play a significant role in cooling CR electrons.  
This is illustrated in Figure \ref{fig:4} where I plot the fractional energy budget for 1.4~GHz emitting CR electrons as a function of magnetic field strength (see caption for assumptions).  
Accordingly, CR electron cooling via processes other than synchrotron emission may become even more efficient if the large scale magnetic fields approach and exceed mG strengths.  
However, other physics may also be at work in such configurations.  
For example, additional synchrotron emission from secondary $e^{\pm}$ arising from $\pi^{0}$ and $\pi^{\pm}$ decay as CR nuclei inelastically scatter of the interstellar gas, may play a role in producing additional synchrotron emission to compensate for these additional energy losses \citep{ejm09c,ltq10}.  
This will not be the case for normal star-forming galaxies.

\section{Conclusions}
I have investigated a simple test which may prove useful for identifying IMF variations in galaxies at high~$z$ (i.e., $z \ga 3$).  
Using a combination of high frequency ($\ga$10~GHz) radio data along with far-infrared and UV data, one can compare radio vs. IR$+$UV SFR estimates which should become increasingly discrepant as the IMF shifts towards one that is top heavy.  
This test makes use of the fact that radio continuum emission becomes dominated by thermal (free-free) emission at higher frequencies, and in addition, that the non-thermal emission of high-$z$ galaxies should become increasingly suppressed due to increased IC losses of CR electrons off of the CMB.  
Thus, deep surveys at frequencies such as the X-band ($8-12$~GHz) should detect the free-free emission from galaxies at $z\ga3$ directly.  
Such tests to identify variations in the IMF of high-$z$ systems should be possible using deep survey imaging from upcoming radio (EVLA, MeerKAT \& SKA/NAA), mm/IR (ALMA/CCAT), and optical/NIR (JWST) facilities.  

\acknowledgements I would like to thank the organizers of the UP2010 conferences for putting together an exciting and diverse program, as well for letting me be a part of it.  

\bibliography{murphy_e}

\end{document}